\documentclass[12pt]{article}
\usepackage{graphics} 
\usepackage{cite}
\usepackage{epsfig}
\newcommand  {\etal}     {{\it et al.}}

\newcommand  {\Bioch}    {{\it Biochemistry\ }}

\newcommand  {\Biopol}   {{\it Biopolymers\ }}

\newcommand  {\EL}       {{\it Europhys.\ Lett.\ }}

\newcommand  {\JACS}     {{\it J.\ Am.\ Chem.\ Soc.\ }}  
\newcommand  {\JBP}      {{\it J.\ Biol.\ Phys.\ \ }}

\newcommand  {\JCP}      {{\it J.\ Chem.\ Phys.\ }}
\newcommand  {\JMB}      {{\it J.\ Mol.\ Biol.\ }}

\newcommand  {\JPC}      {{\it J.\ Phys.\ Chem.\ }}

\newcommand  {\MP}       {{\it Mol.\ Phys.\ }}

\newcommand  {\Nat}      {{\it Nature\ }}
\newcommand  {\NSB}      {{\it Nat.\ Struct.\ Biol.\ }}

\newcommand  {\Pro}      {{\it Proteins\ }}

\newcommand  {\ProSci}   {{\it Protein\ Sci.\ }}

\newcommand  {\PNAS}     {{\it Proc.\ Natl.\ Acad.\ Sci.\ USA\ }}

\newcommand  {\PRL}      {{\it Phys.\ Rev.\ Lett.\ }}

\newcommand  {\Sci}      {{\it Science\ }}

\newcommand{\beq}{\begin{equation}}
\newcommand{\eeq}{\end{equation}}
\newcommand{\beqa}{\begin{eqnarray}}
\newcommand{\eeqa}{\end{eqnarray}}
\newcommand{\bea}{\begin{eqnarray}}
\newcommand{\eea}{\end{eqnarray}}
\newcommand   {\ev}[1]   {\langle #1\rangle}
\newcommand   {\Fs}      {F${}_{\mbox{{\small s}}}$} 

\newcommand   {\Cb}      {C${}_{\beta}$}

\newcommand   {\Cp}      {C${}^{\prime}$}
\newcommand   {\dE}      {\Delta E}
\newcommand   {\dEcal}   {\Delta E_{\mbox{{\scriptsize cal}}}}
\newcommand   {\dEsw}    {\Delta E_{\mbox{{\scriptsize sw}}}}

\newcommand   {\Eev}     {E_{\mbox{{\scriptsize ev}}}}
\newcommand   {\Ehb}     {E_{\mbox{{\scriptsize hb}}}}
\newcommand   {\Ehp}     {E_{\mbox{{\scriptsize hp}}}}
\newcommand   {\Ehpu}    {E_{\mbox{{\scriptsize hp}}}^{\mbox{{\scriptsize u}}}}
\newcommand   {\Ehpf}    {E_{\mbox{{\scriptsize hp}}}^{\mbox{{\scriptsize f}}}}
\newcommand   {\Tm}      {T_{\mbox{{\scriptsize m}}}}
\newcommand   {\Cv}      {C_{\mbox{{\scriptsize v}}}}
\newcommand   {\Of}      {O^{\mbox{{\scriptsize f}}}}
\newcommand   {\Ou}      {O^{\mbox{{\scriptsize u}}}}
\newcommand   {\rc}      {r^{\mbox{{\scriptsize c}}}}

\newcommand   {\eev}     {\epsilon_{\mbox{{\scriptsize ev}}}}

\newcommand   {\ehba}    {\epsilon^{(1)}_{\mbox{{\scriptsize hb}}}}
\newcommand   {\ehbb}    {\epsilon^{(2)}_{\mbox{{\scriptsize hb}}}}
\newcommand   {\ehp}     {\epsilon_{\mbox{{\scriptsize hp}}}}
\newcommand   {\Mij}     {M_{IJ}}

\newcommand   {\shb}     {\sigma_{\mbox{{\scriptsize hb}}}}

\begin{document}

\begin{flushright}
Revised version\\
LU TP 02-28\\
May 20, 2003
\end{flushright}

\vspace{0.4in}

\begin{center}

{\LARGE \bf Thermodynamics of $\alpha$- and $\beta$-Structure 
Formation in Proteins}

\vspace{.6in}

\large
Anders Irb\"ack, Bj\"orn Samuelsson,\\ Fredrik Sjunnesson and Stefan 
Wallin\footnote{E-mail: anders,\,bjorn,\,fredriks,\,stefan@thep.lu.se}\\   
\vspace{0.10in}
Complex Systems Division, Department of Theoretical Physics\\ 
Lund University,  S\"olvegatan 14A,  SE-223 62 Lund, Sweden \\
{\tt http://www.thep.lu.se/complex/}\\

\vspace{0.3in}

\end{center}
\vspace{0.3in}
\normalsize
Abstract:\\
An atomic protein model with a minimalistic potential is developed and
then tested on an $\alpha$-helix and a $\beta$-hairpin, using
exactly the same parameters for both peptides. We find that melting
curves for these sequences to a good approximation can be described 
by a simple two-state model, with parameters that are in
reasonable {\it quantitative} agreement with experimental data.
Despite the apparent two-state character of the melting curves,
the energy distributions are found to lack a clear bimodal shape,
which is discussed in some detail.
We also perform a Monte Carlo-based kinetic study and find,
in accord with experimental data, that the $\alpha$-helix forms
faster than the $\beta$-hairpin.

\newpage

\section{Introduction}

Simulating protein folding at atomic resolution is a challenge, 
but no longer computationally impossible, as shown by recent 
studies~\cite{Shimada:02,Clementi:03} of G\=o-type~\cite{Go:81} models
with a bias towards the native structure. Extending these calculations 
to entirely sequence-based potentials remains, however, an open 
problem, due to well-known uncertainties about the form and  
relevance of different terms of the potential. In this situation, it is 
tempting to look into the properties of atomic models that are 
sequence-based and yet as simple and transparent as possible;
for an example, see Kussell~\etal~\cite{Kussell:02}.

The development of models for protein folding is hampered by the 
fact that short amino acid sequences with protein-like properties
are rare, which makes the calibration of potentials a non-trivial 
task. Breakthrough experiments in the past ten years have, however, 
found examples of such sequences.  
Of particular importance was the discovery of a peptide 
making $\beta$-structure on its own~\cite{Blanco:94}, 
the second $\beta$-hairpin from the protein G B1 domain, 
along with the finding that this 16-amino acid chain, 
like many small proteins, show two-state folding~\cite{Munoz:97}. 
These experiments have stimulated many theoretical studies 
of the folding properties of this sequence, 
including simulations of atomic models with 
relatively detailed semi-empirical potentials~\cite{Dinner:99,Zagrovic:01,
Roccatano:99,Pande:99,Garcia:01,Zhou:01}. Reproducing the melting 
behavior of the $\beta$-hairpin has, however, proven non-trivial, 
as was recently pointed out by Zhou~\etal~\cite{Zhou:01}. 

Here we develop and explore a simple sequence-based atomic model,
which is found to provide a surprisingly good description of the
thermodynamic behavior of this peptide. The same model, with unchanged
parameters, is also applied to an $\alpha$-helical peptide, the
designed so-called \Fs\ peptide with 21 amino
acids~\cite{Lockhart:92,Lockhart:93}. We find that this sequence 
indeed makes an $\alpha$-helix in the model, and our results for
the stability of the helix agree reasonably well with experimental
data~\cite{Lockhart:92,Lockhart:93,Williams:96,Thompson:97}.                   
Finally, we also study Monte Carlo-based kinetics for both
these peptides. Here we investigate the relaxation of ensemble 
averages at the respective melting temperatures. 

\newpage 

\section{Model and Methods}

\subsection{The Model}

Recently, we developed a simple sequence-based model with 5--6 atoms 
per amino acid for helical proteins~\cite{Irback:00,Irback:01,Favrin:02}.
Here we extend that model by incorporating all atoms. The interaction
potential is deliberately kept simple. The chain representation is,
by contrast, detailed; in fact, it is more detailed than in
standard ``all-atom'' models as all hydrogens are explicitly
included. The presence of the hydrogens has the advantage that
local torsion potentials can be avoided. All bond
lengths, bond angles and peptide torsion angles ($180^\circ$) are held fixed,
which means that each amino acid has the Ramachandran torsion
angles $\phi$, $\psi$ and a number of side-chain torsion angles
as its degrees of freedom (for Pro, $\phi$ is held fixed at $-65^\circ$). 
The geometry parameters held constant are derived by statistical
analysis of Protein Data Bank (PDB)~\cite{Bernstein:77} structures.
A complete list of these parameters can be found as supplemental material.

The potential function
\beq
E=\Eev+\Ehb+\Ehp
\label{energy}\eeq
is composed of three terms, representing excluded-volume
effects, hydrogen bonds and effective hydrophobicity forces
(no explicit water), respectively. The remaining part of this
section describes these different terms. Energy parameters are 
quoted in dimensionless units, in which the melting temperature $\Tm$,
defined as the specific heat maximum, is given by  
$k\Tm=0.4462\pm 0.0014$ for the $\beta$-hairpin.
In the next section, the energy scale of the model is set by 
fixing $\Tm$ for this peptide to the experimental midpoint 
temperature, $\Tm=297$\,K~\cite{Munoz:97}.

The excluded-volume energy $\Eev$ is given by 
\beq
\Eev=\eev \sum_{i<j}
\biggl[\frac{\lambda_{ij}(\sigma_i+\sigma_j)}{r_{ij}}\biggr]^{12}\,,
\label{ev}\eeq
where $\eev=0.10$ and $\sigma_i=1.77$, 1.71, 1.64, 1.42 and 1.00\,\AA\ for 
S, C, N, O and H atoms, respectively. Our choice of $\sigma_i$ values 
is guided by the analysis of Tsai~\etal~\cite{Tsai:99}. 
The parameter $\lambda_{ij}$ in Eq.~\ref{ev} reduces
the repulsion between non-local pairs; $\lambda_{ij}=1$ for all pairs 
connected by three covalent bonds
and for HH and OO pairs from adjacent peptide units, and 
$\lambda_{ij}=0.75$ otherwise. The pairs for which $\lambda_{ij}=1$ 
strongly influence the shapes of Ramachandran maps 
and rotamer potentials. The reason for using $\lambda_{ij}<1$ for the 
large majority of all pairs is both computational efficiency and the 
restricted flexibility of chains with only torsional degrees of freedom.
To speed up the calculations, 
the sum in Eq.~\ref{ev} is evaluated using a pair dependent 
cutoff $\rc_{ij}=4.3 \lambda_{ij}$\,\AA. 
   
The hydrogen-bond energy $\Ehb$ has the form 
\beq
\Ehb= \ehba \sum_{j<i-2 \atop {\rm or}\ j>i+1}
u(r_{ij})v(\alpha_{ij},\beta_{ij}) +
 \ehbb \sum u(r_{ij})v(\alpha_{ij},\beta_{ij})\,,
\label{hbonds}\eeq
where $\ehba=3.1$, $\ehbb=2.0$ and the functions $u$ and $v$ are given by 
\begin{eqnarray}
u(r)&=& 5\bigg(\frac{\shb}{r}\bigg)^{12} -
             6\bigg(\frac{\shb}{r}\bigg)^{10}\label{u}\\
v(\alpha,\beta)&=&\left\{
        \begin{array}{ll}
             (\cos\alpha\cos\beta)^{1/2} & 
              \ {\rm if}\ \alpha,\beta>90^{\circ}\label{v}\\
             0  & \ \mbox{otherwise}
        \end{array} \right.
\end{eqnarray}
The first sum in Eq.~\ref{hbonds} represents 
backbone-backbone hydrogen bonds. Term $ij$ in this sum is an 
interaction between the NH and \Cp O groups of amino acids $i$ and 
$j$, respectively.  
$r_{ij}$ denotes the HO distance, and $\alpha_{ij}$ and $\beta_{ij}$ 
are the NHO and HO\Cp\ angles, respectively. The second
sum in Eq.~\ref{hbonds} 
is expressed in a schematic way. It  
represents interactions between oppositely 
charged side chains, and between charged side chains and the 
backbone. Both these types of interaction are, for convenience, taken to 
have the same form as backbone-backbone hydrogen bonds. The side chain 
atoms that can act as ``donors'' or ``acceptors'' in these interactions are 
the N atoms of Lys and Arg (donors) and the O atoms of Asp and Glu 
(acceptors). The second sum in Eq.~\ref{hbonds} has a relatively weak 
influence on the thermodynamic behavior of the systems studied. 
The backbone-backbone hydrogen bonds are, by contrast, crucial
and their strength, $\ehba$, must be carefully chosen~\cite{Irback:01}. 

The functional form of the hydrogen-bond energy differs from
that in our helix model~\cite{Irback:00,Irback:01,Favrin:02}
in that the exponent of the cosines is 1/2 instead of 2. The 
reason for this change is that the $\beta$-hairpin turned out 
to become too regular when using the exponent 2; the exponent 1/2 
gives a more permissive angular dependence. 
The function $u(r)$ in Eq.~\ref{u} is calculated 
using a cutoff $\rc=4.5$\,\AA\ and $\shb=2.0$\,\AA. 

The last term of the potential, the hydrophobicity energy $\Ehp$, 
assigns to each amino acid pair an energy that depends on the amino acid 
types and the degree of contact between the side chains. It can be written as 
\beq
\Ehp=\ehp\sum\Mij C_{IJ}\,,
\label{HP}\eeq
where $\ehp=1.5$, and the sum runs over all possible amino acid pairs $IJ$
except nearest neighbors along the chain. In the present study, the $\Mij$'s 
($\le 0$) are given by the contact energies of Miyazawa and 
Jernigan~\cite{Miyazawa:96} shifted to zero mean, provided that 
the amino acids $I$ and $J$ both are hydrophobic and that the 
shifted contact energy is negative; otherwise, $\Mij=0$. 
The statistical Miyazawa-Jernigan energies contain, 
of course, other contributions too, but receive a major 
contribution from hydrophobicity~\cite{Li:97}. The matrix $\Mij$ 
is given in Table~\ref{tab:1}. Eight of the amino acids are 
classified as hydrophobic, 
namely Ala, Val, Leu, Ile, Phe, Tyr, Trp and Met.  
The geometry factor $C_{IJ}$ in Eq.~\ref{HP} is a measure of the degree
of contact between amino acids $I$ and $J$. To define $C_{IJ}$, we use 
a predetermined set of $N_I$ atoms, denoted by $A_I$, for each amino
acid $I$. For Phe, Tyr and Trp, the set $A_I$ 
consists of the C atoms of the hexagonal ring. The other five 
hydrophobic amino acids each have an $A_I$ containing all its 
non-hydrogen side-chain atoms. With these definitions, $C_{IJ}$
can be written as   
\beq
C_{IJ}=\frac{1}{N_I+N_J}\biggl[\,
\sum_{i\in A_I}f(\min_{j\in A_J} r_{ij}^2) +       
\sum_{j\in A_J}f(\min_{i\in A_I} r_{ij}^2) 
\,\biggr]\,,
\label{Rij}\eeq       
where the function $f(x)=1$ if $x<A$, $f(x)=0$ if $x>B$, 
and $f(x)=(B-x)/(B-A)$ if $A<x<B$ 
[$A=(3.5\,{\rm \AA}){}^2$ and $B=(4.5\,{\rm \AA}){}^2$].
Roughly speaking, $C_{IJ}$
is a measure of the fraction of atoms in $A_I$ or $A_J$ that are in contact 
with some atom from the opposite side chain. 

\begin{table}[t]
\begin{center}
\begin{tabular}{lcccccccc}
    & Ala  & Val  & Leu  & Ile  & Phe  & Tyr  & Trp  & Met  \\
Ala & 0.00 & 0.44 & 1.31 & 0.98 & 1.21 & 0.00 & 0.22 & 0.34 \\
Val &      & 1.92 & 2.88 & 2.45 & 2.69 & 1.02 & 1.58 & 1.72 \\
Leu &      &      & 3.77 & 3.44 & 3.68 & 2.07 & 2.54 & 2.81 \\
Ile &      &      &      & 2.94 & 3.24 & 1.65 & 2.18 & 2.42 \\
Phe &      &      &      &      & 3.66 & 2.06 & 2.56 & 2.96 \\
Tyr &      &      &      &      &      & 0.57 & 1.06 & 1.31 \\
Trp &      &      &      &      &      &      & 1.46 & 1.95 \\
Met &      &      &      &      &      &      &      & 1.86 \\
\end{tabular}
\caption{The interaction matrix $\Mij$, based on the shifted contact-energy 
matrix of Miyazawa and Jernigan~\cite{Miyazawa:96}. 
The table shows absolute values ($\Mij\le0$).} 
\label{tab:1}
\end{center}
\end{table}

\subsection{Numerical Methods}

To study the thermodynamic behavior of this model, we use the   
simulated-tempering method~\cite{Lyubartsev:92,Marinari:92,Irback:95},
in which the temperature is a dynamical variable. This method is chosen
in order to speed up the calculations at low temperatures. Our simulations   
are started from random configurations, and eight different
temperatures are studied, ranging from 273\,K to 366\,K.

The temperature jump is always to a neighboring temperature 
and subject to a Metropolis accept/reject question~\cite{Metropolis:53}. 
For the backbone degrees of freedom, we use three different elementary
moves: first, the pivot move~\cite{Lal:69} in which a single 
torsion angle is turned; second, a semi-local method~\cite{Favrin:01} that 
works with seven or eight adjacent torsion angles, which are turned in a 
coordinated way; and third, a symmetry-based update of 
three randomly chosen backbone torsion angles, 
referred to as the mirror update. 
All updates of side-chain angles and the pivot move are Metropolis 
updates of a single angle, in which the proposed angle is drawn from
the uniform distribution between 0${}^\circ$ and 360${}^\circ$.
To see how the mirror update works, 
consider the three bonds corresponding to the randomly chosen 
torsion angles. The idea is then to reflect the mid bond in the plane 
defined by the two others, keeping the directions of these two other bonds 
fixed. 
Both this update and the pivot move are non-local. They are included
in our thermodynamic calculations in order to accelerate the evolution 
of the system at high temperatures.
The ratio of attempted 
temperature moves to conformation moves is 1:100. 70\% of the conformation 
moves are side-chain moves. The relative ratios of attempts for the three 
types backbone moves is temperature dependent. The 
pivot\,:\,semi-local\,:\,mirror ratio 
varies from 1\,:\,4\,:\,1 at the lowest temperature to 
5\,:\,0\,:\,1 at the highest temperature.
  
Our kinetic simulations are also Monte Carlo-based, and only 
meant to mimic the time evolution of the system in a qualitative 
sense. They differ from our thermodynamic simulations  
in two ways: first, the temperature is held constant; and second,
the two non-local backbone updates are not used, 
but only the semi-local method~\cite{Favrin:01}. 
This restriction is needed 
in order to avoid large unphysical deformations of the chain. 
For the side-chain degrees of freedom, we use a Metropolis 
step in which the angle can change by any amount (same as in the
thermodynamic runs). Thus, it is assumed that the torsion angle 
dynamics are much faster for the side chains than for the backbone.      

In our thermodynamic analysis, statistical errors are obtained by 
analyzing data from ten independent runs, each containing $10^9$ 
elementary steps and several folding/unfolding events. All errors
quoted are 1$\sigma$ errors. 
All fits of data discussed in the next 
section are carried out by using a Levenberg-Marquardt 
procedure~\cite{NR}.      

\section{Results and Discussion}

Using the model described in the previous section, we first study the 
second $\beta$-hairpin from the protein G B1 domain (amino acids 41--56). 
Blanco~\etal~\cite{Blanco:94}
analyzed this peptide in solution by NMR and found that the excised
fragment adopts a structure similar to that in the full protein, although
the NMR restraints were insufficient to determine a unique structure.
In our calculations, in the absence of a complete structure for the 
isolated fragment, we monitor the root-mean-square deviation (rmsd) from 
the native $\beta$-hairpin of the full protein (PDB code 1GB1, first model),  
as determined by NMR~\cite{Gronenborn:91}. The native $\beta$-hairpin 
contains a hydrophobic cluster consisting of Trp43, Tyr45,
Phe52 and Val54. There is experimental evidence~\cite{Kobayashi:00} that
this cluster as well as sequence-specific hydrogen bonds in the turn are
crucial for the stability of the isolated $\beta$-hairpin. 

\begin{figure}
\begin{center}
\epsfig{figure=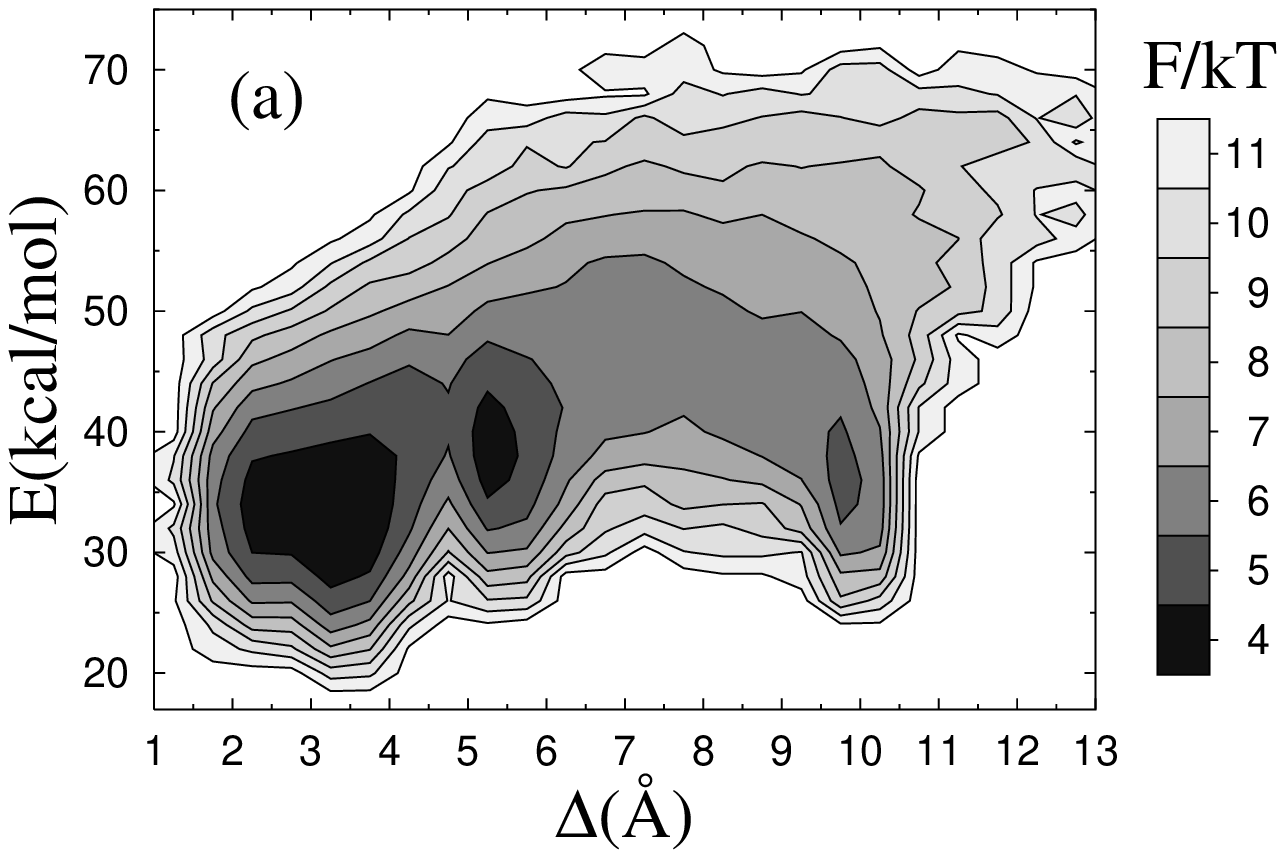,width=7cm}
\hspace{5mm}
\epsfig{figure=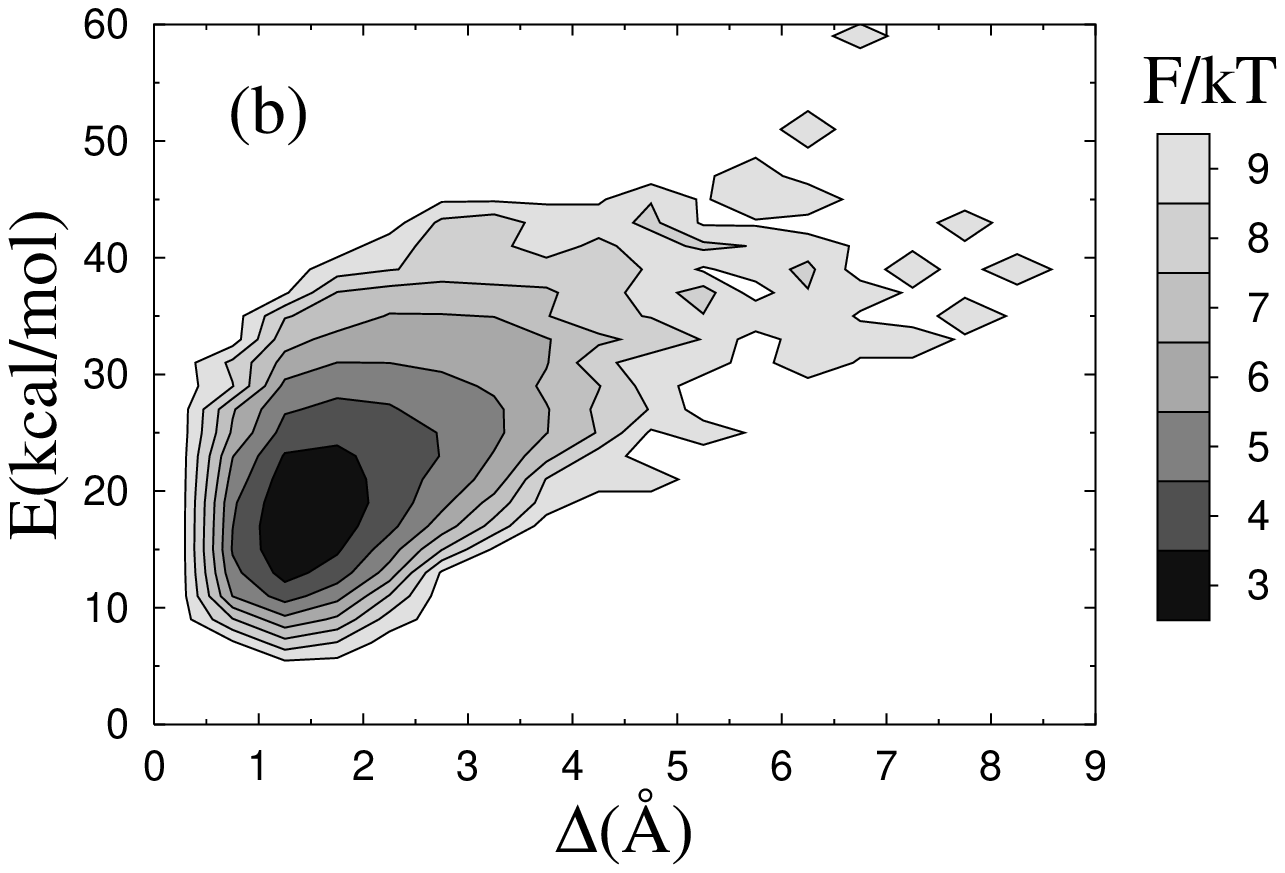,width=7cm}
\end{center}
\caption{Free energy $F(\Delta,E)=-kT\ln P(\Delta,E)$ 
at $T=273$\,K for (a) the $\beta$-hairpin and (b) the \Fs\ peptide.
$E$ is energy and $\Delta$ denotes rmsd from the native 
$\beta$-hairpin
and an ideal $\alpha$-helix, respectively, calculated over all non-hydrogen 
atoms (a backbone rmsd would be unable to distinguish the two possible 
$\beta$-hairpin topologies).}
\label{fig:1}\end{figure}

Fig.~\ref{fig:1}a shows the free energy $F(\Delta,E)$ as a function of rmsd 
from the native $\beta$-hairpin, 
$\Delta$, and energy, $E$, at the temperature $T=273$\,K. 
For a $\beta$-hairpin there are two topologically distinct states with 
similar backbone folds but oppositely oriented side chains.
The global minimum of $F(\Delta,E)$ is found at 2--4\,\AA\ in $\Delta$
and corresponds to a $\beta$-hairpin with the native topology and the 
native set of hydrogen bonds between the two strands. The main difference
between structures within this minimum lies in the shape of the turn. 
The precise shape of the $\beta$-hairpin is, not unexpectedly,  
sensitive to details of the potential; in particular, we find that
the second term in Eq.~\ref{hbonds} does influence the shape of the turn, 
while having only a small effect on thermodynamic functions such as $\Ehp$. 
Therefore, it is not unlikely that a more detailed potential 
would discriminate between different shapes of the turn, and 
thereby make the free-energy minimum more narrow. 

Besides its global minimum, $F(\Delta,E)$ exhibits two local minima
(see Fig.~\ref{fig:1}a), one corresponding to a $\beta$-hairpin
with the non-native topology ($\Delta\approx5$\,\AA),
and the other to an $\alpha$-helix ($\Delta\approx10$\,\AA). A closer
examination of structures from the two $\beta$-hairpin minima
reveals that the \Cb-\Cb\ distances for Tyr45-Phe52 and Trp43-Val54 
tend to be smaller in the non-native topology than in the native one. 
This is important because it makes it sterically
difficult to achieve a proper contact between the aromatic side chains
of Tyr45 and Phe52 in the non-native topology. As a result, this 
topology is hydrophobically disfavored. This is the main reason why the 
model indeed favors the native topology over the non-native one.

\begin{figure}
\begin{center}
\epsfig{figure=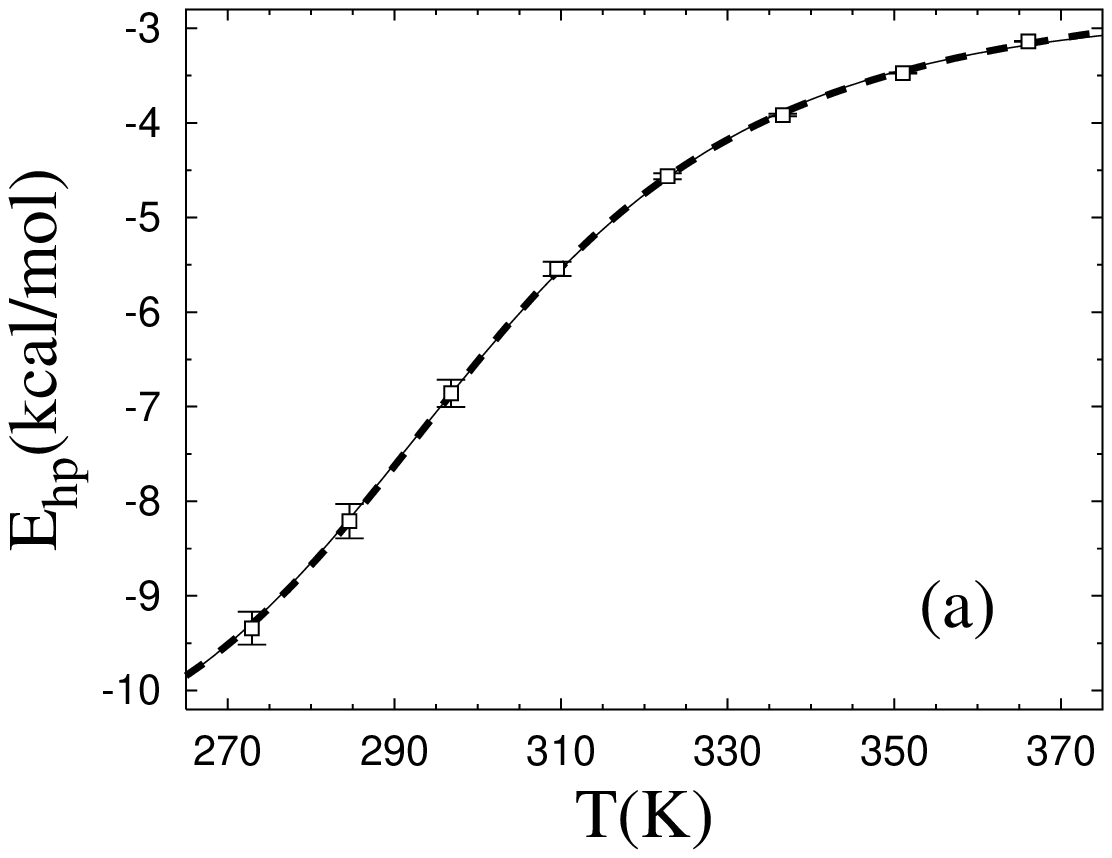,width=7.0cm}
\epsfig{figure=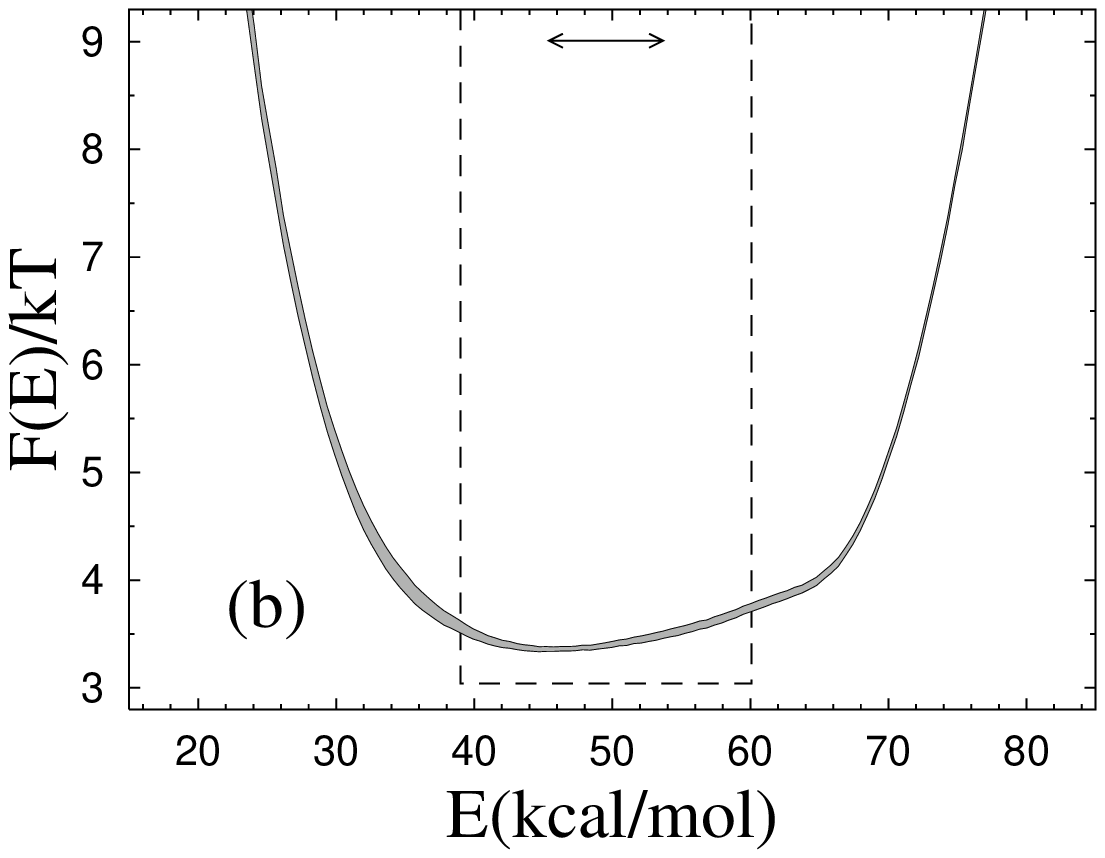,width=7.0cm}
\end{center}
\caption{Unfolding of the $\beta$-hairpin sequence.  (a) Temperature
dependence of the hydrophobicity energy $\Ehp$ (see Eq.~\ref{HP}). 
The solid and dashed curves (essentially coinciding) are fits 
of the data to the two-state expression $\Ehp=(\Ehpu+K\Ehpf)/(1+K)$ 
and the square-well model (see text), respectively. The effective 
equilibrium constant $K$ is assumed to 
have the first-order form $K=\exp[(1/kT-1/k\Tm)\Delta E]$. 
Both fits have three free parameters, whereas 
$\Tm=297$\,K is held fixed.  (b) Free-energy profile $F(E)=-kT\ln P(E)$ at 
$T=\Tm$, obtained by reweighting~\cite{Ferrenberg:88} the data 
at a simulated $T$ close to $\Tm$.   
The shaded band is centered around the expected value and shows 
statistical $1\sigma$ errors. The double-headed arrow indicates $\Delta E$ 
of the two-state fit. The dashed line shows $F(E)$ for 
the square-well fit.}
\label{fig:2}
\end{figure}

We now turn to the melting behavior of the $\beta$-hairpin. By studying 
tryptophan fluorescence (Trp43), Mu\~noz~\etal~\cite{Munoz:97} found that the
unfolding of this peptide with increasing temperature shows two-state
character, with parameters $\Tm=297$\,K and $\Delta E=11.6$\,kcal/mol,
$\Tm$ and $\Delta E$ being the melting temperature and energy change, 
respectively. To study the character of the melting transition in our model, 
we monitor the hydrophobicity energy $\Ehp$, a simple observable we 
expect to be strongly correlated with Trp43 fluorescence. 
Following Mu\~noz~\etal~\cite{Munoz:97}, we fit our data for $\Ehp$ 
to a first-order two-state model. To reduce the number
of parameters of the fit, $\Tm$ is held fixed, at the specific 
heat maximum (data not shown). The fit turns out not to be perfect, with 
a $\chi^2/{\rm dof}$ of 4.5. The deviations from the fitted curve are
nevertheless small, as can be seen from Fig.~\ref{fig:2}a; 
they can be detected only because the statistical errors 
are very small ($\sim 0.1$\,\%) at the highest temperatures.
To further illustrate this point, we assign each data point 
an artifical uncertainty of 1\,\%, an error size that is not uncommon    
for experimental data. With these errors, the same type of fit yields 
a $\chi^2/$dof of 0.3, which confirms that the data indeed to a 
good approximation show two-state behavior. Our fitted value of $\Delta E$ 
is $9.3\pm0.3$\,kcal/mol, which implies that the temperature dependence 
of the model is comparable to experimental data~\cite{Munoz:97}. 

Several groups have simulated the same $\beta$-hairpin using atomic models
with implicit~\cite{Dinner:99,Zagrovic:01,Kussell:02} or 
explicit~\cite{Roccatano:99,Pande:99,Garcia:01,Zhou:01} solvent. 
Many of these groups studied the melting behavior of the $\beta$-hairpin,
but the temperature dependence they found was too weak, 
as was pointed out by Zhou~\etal~\cite{Zhou:01}.
In fact, in all these studies, there was a significant $\beta$-hairpin 
population at temperatures of 400\,K and above.
Another important difference between at least some of these 
models~\cite{Zagrovic:01,Pande:99,Garcia:01} and ours, is that in our 
model there is no clear free-energy minimum corresponding to a 
hydrophobically collapsed state with few or no hydrogen bonds. 
A local free-energy minimum with helical content was found in one 
of these studies~\cite{Garcia:01}, but not in the others. Such a   
minimum exists in our model (see Fig.~\ref{fig:1}a), but the helix 
population is low. 

\begin{figure}
\begin{center}
\epsfig{figure=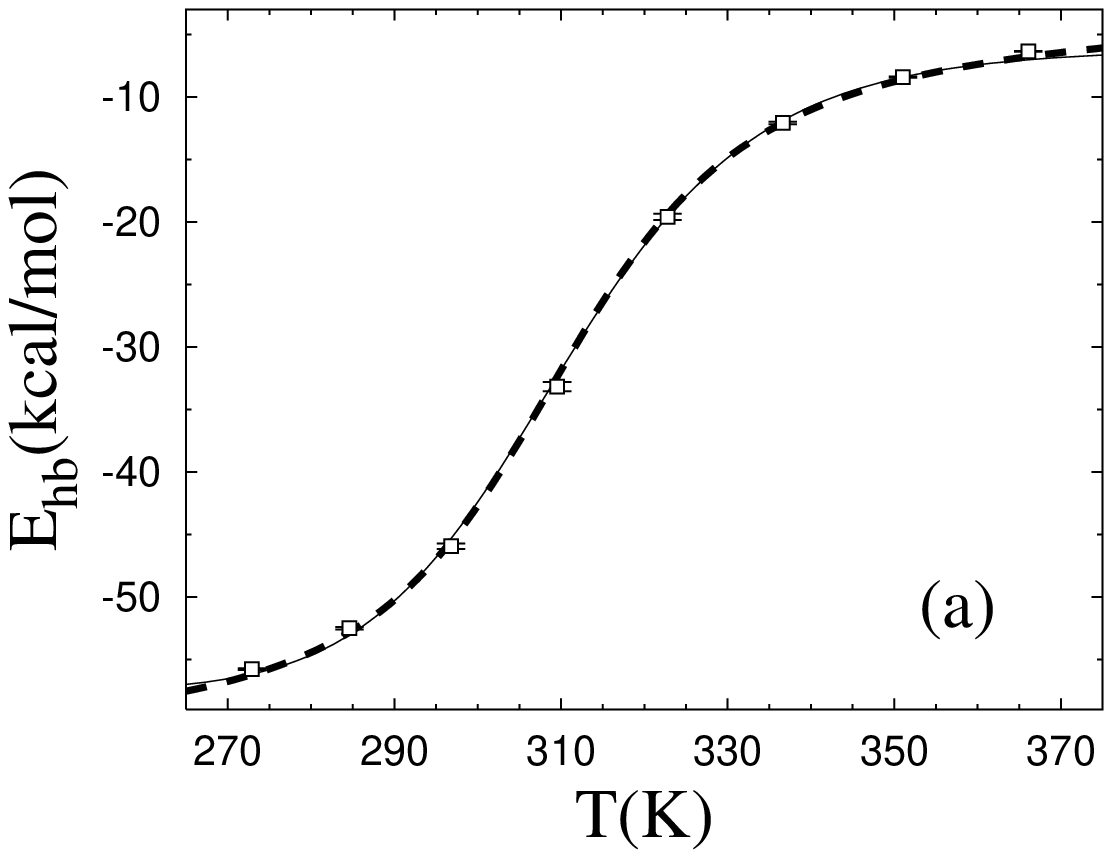,width=7.0cm}
\epsfig{figure=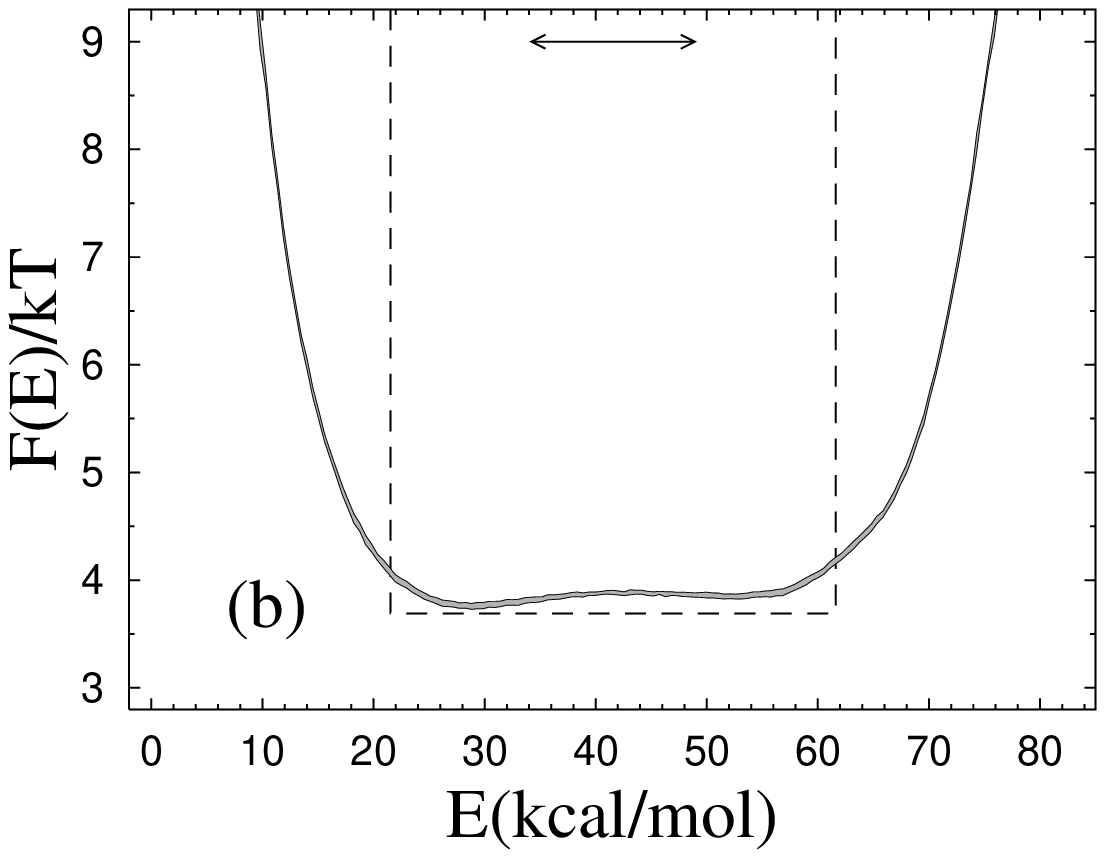,width=7.0cm}
\end{center}
\caption{Unfolding of the \Fs\ sequence. (a) Temperature dependence
of the hydrogen-bond energy $\Ehb$ (see Eq.~\ref{hbonds}), with the same 
two types of fit as in Fig.~\ref{fig:2}a (same symbols).  
(b) Free-energy profile $F(E)=-kT\ln P(E)$ at $T=\Tm$. Same symbols as
in Fig.~\ref{fig:2}b.}
\label{fig:3}
\end{figure}

In spite of its minimalistic potential, our model  
is able to make $\alpha$-helices too. To show this, 
we consider the $\alpha$-helical so-called 
\Fs\ peptide, which has been extensively studied both
experimentally~\cite{Lockhart:92,Lockhart:93,Williams:96,Thompson:97}
and theoretically~\cite{Garcia:02}. This 21-amino acid peptide is
given by AAAAA(AAARA)${}_3$A, where A is Ala and R is Arg.
Using exactly the same model as before, with unchanged parameters,
we find that the \Fs\ sequence does make an $\alpha$-helix.
This can be seen from Fig.~\ref{fig:1}b, which shows the
free energy $F(\Delta,E)$ at $T=273$\,K, $\Delta$ 
this time denoting rmsd from an ideal $\alpha$-helix.      
$F(\Delta,E)$ has only one significant minimum, which 
indeed is helical.  
The melting behavior of this sequence is illustrated in Fig.~\ref{fig:3}a, 
which shows the temperature dependence of the hydrogen-bond energy. 
Data are again quite well described by a first-order two-state model; 
the $\chi^2/{\rm dof}$ for the fit is 20.5 and would be 1.7 if the errors 
were 1\,\%. Our fitted value of $\Delta E$ is $16.1\pm0.9$\,kcal/mol 
for \Fs, which may be compared to the result 
$\dE=12\pm2$\,kcal/mol obtained by a two-state fit of infrared (IR) 
spectroscopy data~\cite{Williams:96}. As in the $\beta$-hairpin analysis, 
$\Tm$ is determined from the specific heat maximum (data not shown). For
\Fs, we obtain $\Tm=310$\,K, which may be compared 
to the values $\Tm=303$, 308\,K and $\Tm=334$\,K
obtained by circular dichroism (CD)~\cite{Lockhart:93,Thompson:97} and
IR spectroscopy~\cite{Williams:96}, respectively.    
Let us stress that $\Tm$ for \Fs\ is
a prediction of the model; the energy scale of the model is set using $\Tm$ 
for the $\beta$-hairpin and then left unchanged in our study of 
\Fs. 

The two-state fits shown in Figs.~\ref{fig:2}a and \ref{fig:3}a are based 
on a first-order expression for the free energies of the two coexisting 
phases. The fits look good and can be improved by including higher 
order terms, which may give the impression that the behaviors of these 
systems can be fully understood in terms of a two-state model.
However, the two-state picture is far from perfect. 
This can be seen from the free-energy profiles $F(E)$ 
shown in Figs.~\ref{fig:2}b and \ref{fig:3}b, which lack a clear 
bimodal shape. Clearly, this renders the parameters of a two-state 
model, such as $\Delta E$, ambiguous. The analysis of these
systems therefore shows that the results of a two-state fit must be 
interpreted with care. Given the actual shapes of $F(E)$, 
it is instructive to perform an alternative fit of the data 
in Figs.~\ref{fig:2}a and \ref{fig:3}a, based on the assumptions that 
1) $F(E)$ has the shape of a square well of width $\dEsw$ at $T=\Tm$, 
and that 2) the observable analyzed varies linearly with $E$.\footnote{
With these two assumptions, one finds that the average value of an
arbitrary observable $O$ at temperature $T$ is given by 
$O(T)=\int_0^1(\Ou(1-t)+\Of t)\lambda^tdt\Big/\int_0^1\lambda^tdt= 
\Ou+(\Of-\Ou)(\frac{\lambda}{\lambda-1}-\frac{1}{\ln \lambda})$, 
where $\lambda=\exp[(1/kT-1/k\Tm)\dEsw]$
and $\Ou$ and $\Of$ are the values of $O$ at the respective edges of 
the square well.} 
These square-well fits are shown in Figs.~\ref{fig:2}a
and \ref{fig:3}a, and the corresponding free-energy profiles $F(E)$
(at $T=\Tm$) are indicated in Figs.~\ref{fig:2}b and 
\ref{fig:3}b. 
The square-well fits are somewhat better than the two-state 
fits. However, the fitted curves are strikingly similar, 
given the large difference between the underlying energy distributions. 
This shows that it is very hard to draw conclusions about the 
free-energy profile $F(E)$ from the temperature dependence of a 
single observable. 

From Figs.~\ref{fig:2}b and \ref{fig:3}b it can also be seen that the 
energy change $\dE$ obtained from the two-state fit is considerably
smaller than the width of the energy distribution, which indicates that
$\dE$ is smaller than the calorimetric energy change $\dEcal$.
Scholtz~\etal~\cite{Scholtz:91} determined $\dEcal$ experimentally for
an Ala-based helical peptide with 50 amino acids, and obtained a 
value of 1.3\,kcal/mol per amino acid. This value corresponds to a $\dEcal$ 
of 27.3 kcal/mol for the \Fs\ peptide. Comparing model results for 
$\dEcal$ with experimental data is not straightforward, due to 
uncertainties about what the relevant baseline subtractions
are~\cite{Zhou:99,Chan:00,Kaya:00}. If we ignore baseline subtractions
and simply define $\dEcal$ as the energy change between the
highest and lowest temperatures studied, we obtain
$\dEcal=45.6\pm0.1$\,kcal/mol for \Fs, which is larger 
than the value of Scholtz~\etal~\cite{Scholtz:91}. To get an idea of
how much this result can be affected by a baseline subtraction, a
fit of our specific heat data is performed, to a two-state expression 
supplemented with a baseline linear in $T$. The fit function is 
$\Cv=\dEcal(1+K)^{-2}\frac{dK}{dT}+c_0+c_1(T-\Tm)$, where
$c_0$ and $c_1$ are baseline parameters and
$K=\exp[(1/kT-1/k\Tm)\dE]$.  With $\dEcal$, $\dE$, $c_0$, $c_1$ and
$\Tm$ as free parameters, this fit gives $\dEcal=34.0\pm1.0$\,kcal/mol
($\chi^2/{\rm dof}=5.2$), which is considerably closer to the   
value of Scholtz~\etal~\cite{Scholtz:91}. It may be worth noting that 
the corresponding fit without baseline subtraction is much poorer 
($\chi^2/{\rm dof}\sim 300$).  From these calculations, we conclude
that the model may overestimate $\dEcal$, but it is not evident that   
the deviation is statistically significant, due to  
theoretical as well as experimental uncertainties. 

The melting behavior of helical peptides is often analyzed using the 
Zimm-Bragg~\cite{Zimm:59} or Lifson-Roig~\cite{Lifson:60} models, 
which for large chain lengths are very different from the two-state 
model considered above. Our results for the \Fs\ peptide are,
nevertheless, quite well described by these models too. 
In fact, a fit of the helix content as a function of temperature 
to the Lifson-Roig model gives a $\chi^2/{\rm dof}$ similar to 
that for the two-state
fit above.\footnote{We define helix
content in the following way. Each amino acid, except the two at the ends, 
is labeled h if $-90^\circ<\phi<-30^\circ$ and $-77^\circ<\psi<-17^\circ$, 
and c otherwise. $j$ consecutive h's form a helical segment of length 
$j-2$. The maximal number of amino acids in helical segments is then
$N-4$ for a chain with $N$ amino acids.} 
Our fitted Lifson-Roig parameters are $v=0.016\pm0.009$  
and $w(T=273\,{\rm K})=1.86\pm0.25$, 
corresponding to the Zimm-Bragg parameters $\sigma=0.0003\pm0.0003$ and
$s(T=273\,{\rm K})=1.83\pm0.25$~\cite{Qian:92}. 
In this fit the temperature dependence of $w$ is given by a first-order
two-state expression, whereas $v$ is held constant. The energy change  
$\Delta E_w$ has a fitted value of $1.33\pm0.17$\,kcal/mol. The
statistical uncertainties on $v$ and $\sigma$ are large because the 
chain is small, which makes the dependence on these parameters weak.
Thompson~\etal~\cite{Thompson:97} performed a Zimm-Bragg analysis
of CD data for \Fs, using the single-sequence approximation.
Assuming a value of $\Delta E_s=1.3$\,kcal/mol for the energy change 
associated with helix propagation, they obtained a $\sigma$ of 0.0012.

\begin{figure}
\begin{center}
\epsfig{figure=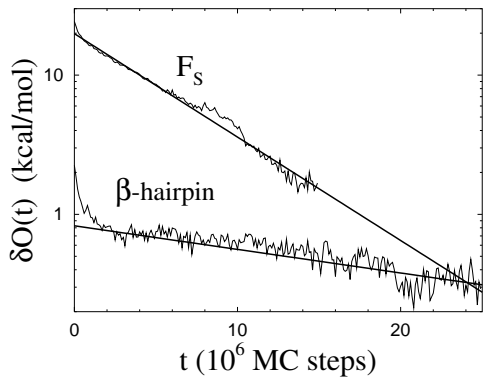,width=7.0cm}
\end{center}
\caption{Monte Carlo relaxation of ensemble averages at $T=\Tm$ for
  the $\beta$-hairpin and the \Fs\ peptide.
  The deviation $\delta O(t)$ from the
  equilibrium average (see text) is plotted against 
  the number of elementary Monte Carlo steps, $t$. 
  Straight lines are $\chi^2$ fits of 
  the data to a single exponential. Data for $t>15\cdot 10^6$ 
  are omitted for \Fs\ due to large statistical errors.}
\label{fig:4}
\end{figure}

Our kinetic simulations of the two peptides are performed at their
respective melting temperatures, $\Tm$. Starting from equilibrium
conformations at $T=366$\,K, we study the relaxation of ensemble
averages under Monte Carlo dynamics (see Section 2.2). The ensemble
consists of 1500 independent runs for each peptide.
In Fig.~\ref{fig:4}, we show the ``time'' evolution of $\delta
O(t)=O(t)-\ev{O}$, where $O(t)$ is an ensemble average after $t$ Monte
Carlo steps, $\ev{O}$ is the corresponding equilibrium average,  
and the observable $O$ is $\Ehp$ for the $\beta$-hairpin and 
$\Ehb$ for \Fs\ (same observables as in the thermodynamic calculations).  
Ignoring a brief initial period of rapid change, 
we find that the data, for both peptides, are fully consistent
with single-exponential relaxation ($\chi^2/{\rm dof}\sim1$), although 
the interval over which the signal $\delta O(t)$ can be
followed is small in units of the relaxation time, especially 
for the $\beta$-hairpin. Nevertheless, assuming the single-exponential 
behavior to be correct, a statistically quite accurate determination 
of the relaxation times can be obtained. The fitted relaxation time 
is approximately a factor of 5 larger for the $\beta$-hairpin 
than for \Fs. The corresponding factor is around
30 for experimental data~\cite{Munoz:97,Williams:96,Thompson:97}.
A closer look at the $\beta$-hairpin data shows that the hydrophobic
cluster and the hydrogen bonds, on average, form 
nearly simultaneously in our model. This is in agreement with
the results of Zhou~\etal~\cite{Zhou:01}, and in disagreement with
the folding mechanism of Pande and Rokhsar~\cite{Pande:99} in which
the collapse occurs before the hydrogen bonds form. 

The two peptides studied in this paper make unusually clear-cut 
$\alpha$- and $\beta$-structure, respectively. It is clear 
that refinements of the interaction potential will be required  
in order to obtain an equally good description of more general sequences. 
One interesting refinement would be to make the strength of 
the hydrogen bonds context-dependent, that is dependent on 
whether the hydrogen bond is internal or exposed. This is 
probably needed in order for the model to capture, for example, the 
difference between the Ala-based \Fs\ peptide and pure polyalanine. 
In fact, it has been argued~\cite{Garcia:02,Vila:00} 
that a major reason why \Fs\ is a strong helix maker is that the 
Arg side chains shield the backbone from water and thereby make 
the hydrogen bonds stronger. The hydrogen bonds of a 
polyalanine helix lack this protection. In our model, the hydrogen bonds
are context-independent, which could make polyalanine too helical.
Although a direct  
comparison with experimental data is impossible due to its poor water 
solubility, simulations of polyalanine with 21 amino acids, A${}_{21}$, 
seem to confirm this. For A${}_{21}$, we obtain a helix content of 
about 80\,\% at $T=273$\,K, which is what we find for \Fs\ too.
Using a modified version of the Cornell~\etal\ force field~\cite{Cornell:95}, 
Garc\'\i a and Sanbonmatsu~\cite{Garcia:02} obtained a helix content
of 34\,\% at $T=275$\,K for A${}_{21}$; the unmodified force field 
was found~\cite{Garcia:02} to give a value similar to ours at this 
temperature (but very different from ours at higher $T$).  
Our estimate that \Fs\ is $\sim$\,80\,\% helical 
at $T=273$\,K is consistent with experimental 
data~\cite{Lockhart:92,Thompson:97}. 

We also looked at two other helical peptides.
The first of these is the Ala-based 16-amino acid peptide 
(AEAAK)${}_3$A, where E is Glu 
and K is Lys. By CD, Marqusee and Baldwin~\cite{Marqusee:87} found 
this peptide to be $\sim$\,50\,\% helical  
at $T=274$\,K. In our model the corresponding value 
turns out to be $\sim$\,70\,\%. Our last example is   
the 38--59-fragment of the B domain of staphylococcal protein A
(PDB code 1BDD). This is a more general, not Ala-based sequence, containing
three hydrophobic Leu. 
By CD, Bai~\etal~\cite{Bai:97} obtained a helix content of 
$\sim$\,30\,\% at pH 5.2 and $T=278$\,K for this fragment. In our model   
we obtain a helix content of $\sim$\,20\,\% at this temperature. 
So, the model predicts helix contents that are in approximate agreement 
with experimental data for \Fs, (AEAAK)${}_3$A as well as 
the protein A fragment.        

\section{Summary and Outlook}
 
We have developed and explored a protein model that 
combines an all-atom representation of the amino acid chain with a 
minimalistic sequence-based potential. The strength of the model
is the simplicity of the potential, which at the same time, of course, 
means that there are many interesting features of real proteins that 
the model is unable to capture. One advantage of the model is that the 
calibration of parameters, which any model needs, becomes easier to
carry out with fewer parameters to tune. 

When calibrating the model, our goal was to ensure that, without 
resorting to parameter changes, our two sequences  
made a $\beta$-hairpin with the native topology and an 
$\alpha$-helix, respectively, which was not an easy task. 
Once this goal had been achieved, 
our thermodynamic and kinetic measurements 
were carried out without any further fine-tuning of the potential. 
Therefore, it is hard to believe that the generally quite good agreement 
between our thermodynamic results and experimental data is accidental. 
A more plausible
explanation of the agreement is that the thermodynamics of these two 
sequences indeed are largely governed by backbone hydrogen bonding 
and hydrophobic collapse forces, as assumed by the model. The requirement
that the two sequences make the desired structures is then sufficient  
to quite accurately determine the strengths of these two terms. 

The main results of our calculations can be summarized as follows.

\begin{itemize}

\item Our thermodynamic simulations show first of all that the two sequences 
studied indeed make a $\beta$-hairpin with the native topology and an 
$\alpha$-helix, respectively. The main reason why the model favors the 
native topology over the non-native one for the $\beta$-hairpin, 
is that the formation of the hydrophobic cluster is sterically 
difficult to accomplish in the non-native topology. 
The melting curves obtained for the two peptides are 
in reasonable agreement with experimental data, and can to  
a good approximation be described by a simple two-state model.

\item A two-state description of the thermodynamic behavior is, nevertheless, 
found to be an oversimplification for both peptides, as can be seen from the 
energy distributions. Given that the systems are small and fluctuations 
therefore relatively large, this is maybe not surprising. What is striking is 
how difficult it is to detect these deviations from two-state behavior 
when studying the temperature dependence of a single observable.
  
\item  The results of our Monte Carlo-based kinetic runs at the respective 
melting temperatures are, for both peptides, consistent with 
single-exponential relaxation, and the relaxation time is found 
to be larger for the $\beta$-hairpin than for \Fs.  

\end{itemize}

\newpage

Extending these calculations to larger chains will impose new
conditions on the interaction potential, and thereby make it possible
(and necessary) to refine it. Two interesting refinements would be
to make the treatment of charged side chains and side-chain
hydrogen bonds less crude and to introduce a mechanism for screening
of hydrogen bonds~\cite{Garcia:02,Takada:99,Vila:00,Guo:02}. 
Computationally, there is room for extending the calculations.  
In fact, simulating the thermodynamics of a chain with about 20 amino 
acids, with high statistics, does not take more than a few days on a 
standard desktop computer, in spite of the detailed geometry of the model. 
This gives us hope to be able to look into the free-energy landscape 
and two-state character of small proteins in a not too distant future.

\subsection*{Acknowledgments}

We thank Giorgio Favrin for stimulating discussions and help with
computers. 
This work was in part supported by the Swedish Foundation for Strategic 
Research and the Swedish Research Council.


\newpage

\begin{thebibliography}{}

\bibitem{Shimada:02}
Shimada, J. \& Shakhnovich, E.I.
(2002) \PNAS {\bf 99}, 11175--11180.
 
\bibitem{Clementi:03}
Clementi, C., Garc\'\i a, A.E. \& Onuchic, J.N.
(2003) \JMB {\bf 326}, 933--954. 

\bibitem{Go:81}
G\=o, N. \& Abe, H.
(1981) \Biopol {\bf 20}, 991--1011.

\bibitem{Kussell:02}
Kussell, E., Shimada, J. \& Shakhnovich, E.I. 
(2002) \PNAS {\bf 99}, 5343--5348.

\bibitem{Blanco:94}
Blanco, F.J., Rivas, G. \& Serrano, L.
(1994) \NSB {\bf 1}, 584--590.

\bibitem{Munoz:97}
Mu\~noz, V., Thompson, P.A., Hofrichter, J. \& Eaton, W.A.
(1997) \Nat {\bf 390}, 196--199.



\bibitem{Dinner:99}
Dinner, A.R., Lazaridis, T. \& Karplus, M.
(1999) \PNAS {\bf 96}, 9068--9073.

\bibitem{Zagrovic:01}
Zagrovic, B., Sorin, E.J. \& Pande, V.
(2001) \JMB {\bf 313}, 151--169.

\bibitem{Roccatano:99}
Roccatano, D., Amadei, A., Di Nola, A. \& Berendsen, H.J.C.
(1999) \ProSci {\bf 8}, 2130--2143.

\bibitem{Pande:99}
Pande, V.S. \& Rokhsar, D.S.
(1999) \PNAS {\bf 96}, 9062--9067.

\bibitem{Garcia:01}
Garc\'\i a, A.E.  \&  Sanbonmatsu, K.Y.
(2001) \Pro {\bf 42}, 345--354.

\bibitem{Zhou:01}
Zhou, R., Berne, B.J. \& Germain, R.
(2001) \PNAS {\bf 98}, 14931--14936.

\bibitem{Lockhart:92}
Lockhart, D.J. \& Kim, P.S.
(1992) \Sci {\bf 257}, 947--951.
 
\bibitem{Lockhart:93}
Lockhart, D.J. \& Kim, P.S.
(1993) \Sci {\bf 260}, 198--202.

\bibitem{Williams:96}
Williams, S., Causgrove, T.P., Gilmanshin, R., Fang, K.S., Callender, R.H.,
Woodruff, W.H. \& Dyer, R.B.
(1996) \Bioch {\bf 35}, 691--697.

\bibitem{Thompson:97}
Thompson, P.A., Eaton, W.A. \& Hofrichter, J.
(1997) \Bioch {\bf 36}, 9200--9210.

\bibitem{Irback:00}
Irb\"ack, A., Sjunnesson, F. \& Wallin, S.
(2000) \PNAS {\bf 97}, 13614--13618.

\bibitem{Irback:01}
Irb\"ack, A., Sjunnesson, F. \& Wallin, S.
(2001) \JBP {\bf 27}, 169--179.

\bibitem{Favrin:02}
Favrin, G., Irb\"ack, A. \& Wallin, S.
(2002) \Pro {\bf 47}, 99--105.

\bibitem{Bernstein:77}
Bernstein, F.C., Koetzle, T.F., Williams, G.J.B., Meyer, E.F.,
Brice, M.D., Rodgers, J.R., Kennard, O., Shimanouchi, T. \&
Tasumi, M.
(1977) \JMB {\bf 112}, 535--542.

\bibitem{Tsai:99}
Tsai, J., Taylor, R., Chothia, C. \& Gerstein, M.
(1999) \JMB {\bf 290}, 253--266.

\bibitem{Miyazawa:96}
Miyazawa, S. \& Jernigan, R.L.
(1996) \JMB {\bf 256}, 623--644.

\bibitem{Li:97}
Li, H., Tang, C. \& Wingreen, N.S.
(1997) \PRL {\bf 79}, 765--768.

\bibitem{Lyubartsev:92}
Lyubartsev, A.P., Martsinovski, A.A., Shevkunov, S.V. \& 
Vorontsov-Velyaminov, P.N.
(1992) \JCP {\bf 96}, 1776--1783.

\bibitem{Marinari:92}
Marinari, E. \& Parisi, G.
(1992) \EL {\bf 19}, 451--458.

\bibitem{Irback:95}
Irb\"ack, A. \& Potthast, F.
(1995) \JCP {\bf 103}, 10298--10305.

\bibitem{Metropolis:53}
Metropolis, N., Rosenbluth, A.W., Rosenbluth, M.N., Teller, A.H.
\& Teller, E.
(1953) \JCP {\bf 21}, 1087--1092.

\bibitem{Lal:69}
Lal, M. 
(1969) \MP {\bf 17}, 57--64. 

\bibitem{Favrin:01}
Favrin, G., Irb\"ack, A. \&  Sjunnesson, F.
(2001) \JCP {\bf 114}, 8154--8158.

\bibitem{NR}
Press, W.H., Flannery, B.P., Teukolsky, S.A. \& Vetterling, W.T.
(1992) {\it Numerical Recipes in C: The Art of Scientific Computing},
(Cambridge University Press, Cambridge). 

\bibitem{Gronenborn:91}
Gronenborn, A.M., Filpula, D.R., Essig, N.Z., Achari, A.,
Whitlow, M., Wingfield, P.T. \& Clore, G.M.
(1991) \Sci {\bf 253}, 657--661.

\bibitem{Kobayashi:00}
Kobayashi, N., Honda, S., Yoshii, H. \& Munekata, E.
(2000) \Bioch {\bf 39}, 6564--6571.

\bibitem{Garcia:02}
Garc\'\i a, A.E. \&  Sanbonmatsu, K.Y.
(2002) \PNAS {\bf 99}, 2782--2787.

\bibitem{Ferrenberg:88}
Ferrenberg, A.M. \& Swendsen R.H.
(1988) \PRL {\bf 61}, 2635--2638, and erratum (1989) {\bf 63}, 1658,
and references given in the erratum.

\bibitem{Scholtz:91}
Scholtz, J.M., Marqusee, S., Baldwin, R.L., York, E.J., Stewart, J.M.,
Santaro, M. \& Bolen, D.W.
(1991) \PNAS {\bf 88}, 2854--2858.

\bibitem{Zhou:99}
Zhou, Y., Hall, C.K. \& Karplus, M.
(1999) \ProSci {\bf 8}, 1064--1074.

\bibitem{Chan:00}
Chan, H.S.
(2000) \Pro {\bf 40}, 543--571.

\bibitem{Kaya:00}
Kaya, H. \& Chan, H.S.
(2000) \Pro {\bf 40}, 637--661.

\bibitem{Zimm:59}
Zimm, B.H. \& Bragg, J.K.
(1959) \JCP {\bf 31}, 526--535.

\bibitem{Lifson:60}
Lifson, S. \& Roig, A.
(1960) \JCP {\bf 34}, 1963--1974.

\bibitem{Qian:92}
Qian, H. \& Schellman, J.A.
(1992) \JPC {\bf 96}, 3987--3994.

\bibitem{Vila:00}
Vila, J.A., Ripoll, D.R. \& Scheraga, H.A.
(2000) \PNAS {\bf 97}, 13075--13079.

\bibitem{Cornell:95}
Cornell, W.D, Cieplak, P, Bayly, C.I., Gould, I.R, Merz, K.M., Ferguson, D.M.,
Spellmeyer, D.C., Fox, T., Caldwell, J.W. \& Kollman, P.A.
(1995) \JACS {\bf 117}, 5179--5197. 

\bibitem{Marqusee:87}
Marqusee, S. \& Baldwin, R.L.
(1987) \PNAS {\bf 84}, 8898--8902.

\bibitem{Bai:97}
Bai, Y., Karimi, A., Dyson, H.J. \& Wright, P.E.
(1997) \ProSci {\bf 6}, 1449--1457.

\bibitem{Takada:99}
Takada, S., Luthey-Schulten, Z. \& Wolynes, P.G. 
(1999) \JCP {\bf 110}, 11616--11629.

\bibitem{Guo:02}
Guo, C., Cheung, M.S., Levine, H. \& Kessler, D.A.
(2002) \JCP {\bf 116}, 4353--4365. 

\end{thebibliography}
\end{document}